\documentclass[11pt,a4paper]{article}

\usepackage{epsfig,amsmath,amssymb,cite}

\begin{document}

\title{Regular scale-dependent black holes as gravitational lenses}
\author{Carlos M. Sendra$^{1,2,}$\thanks{e-mail: cmsendra@iafe.uba.ar} \\
{\small $^1$ Instituto de Astronom\'{\i}a y F\'{\i}sica del Espacio (IAFE, CONICET-UBA),} \\
{\small Casilla de Correo 67, Sucursal 28, 1428, Buenos Aires, Argentina}\\
{\small $^2$ Departamento de F\'{\i}sica, Facultad de Ciencias Exactas y Naturales, Universidad de Buenos Aires, } \\
{\small Ciudad Universitaria Pabell\'on I, 1428, Buenos Aires, Argentina} }
\maketitle
\date{}

\begin{abstract}
In this article, we study regular scale-dependent black holes as gravitational lenses. We obtain the deflection angle in the strong deflection limit, from which we calculate the positions and the magnifications of the relativistic images. We also compare our results with those corresponding to the Schwarzschild spacetime, and as an example, we apply and discuss these results to the particular case of the supermassive black hole at the center of our galaxy.
\end{abstract}

\section{Introduction}\label{intro}
Gravitational lensing by black holes has received a growing interest since the discovery of supermassive black holes at the center of most galaxies, included ours \cite{gillessen17}. Within this context, the first direct image of the supermassive black hole M87* at the center of the galaxy M87 has been recently obtained \cite{eth1,eth2,eth3,eth4,eth5,eth6} and it is expected that some other optical effects will also be observed in the near future \cite{observ}. In the case of compact objects with a photon sphere, for a point source there exist two infinite sets of the denominated relativistic images \cite{virbha1}, as a consequence of light rays passing close to the photon sphere, besides the primary and secondary images. In this case, the deflection angle results larger than $2\pi$, and can be treated analytically by performing the strong deflection limit, which consists in a logarithmic approximation of this angle. This method allows to obtain analytical expressions for the positions and the magnifications of the relativistic images. The strong deflection limit was firstly introduced for the Schwarzschild geometry \cite{darwin-otros}, extended to the Reissner-Nordstr\"om black hole \cite{eiroto}, and generalized to any spherically symmetric and asymptotically flat spacetime \cite{bozza02,tsukamoto}. Many works related to lenses in the strong deflection limit have been considered in the recent years \cite{nakedsing1,nakedsing2,retro,lenseq,virbha2,alternative1, alternative2,bwlens1,bwlens2,scalarlens,cm, tsukamoto-wh, liang}, both in the context of general relativity and in alternative theories of gravity. The lensing effects for rotating geometries were also studied \cite{rotbh}, including the analysis of the deformation of the shadow as well \cite{shadows1,shadows11,shadows2,shadows22}.

In general relativity, the presence of singularities is a long standing problem, since they are points in spacetime where the theory breaks down. In opposition, regular black hole solutions (i.e., singularity-free) have been obtained, where the event horizon is present but without singularities in its interior. The first regular black hole solution was introduced by Bardeen in 1968 \cite{bardeen} and many other solutions were found employing nonlinear electrodynamics as the matter source (see Ref. \cite{sadaji} and references therein).

In the context of quantum gravity, quantum corrections to black hole solutions can be considered by taking into account an effective action where the coupling constants are replaced by scale-dependent quantities (i.e. $\{G_0,\Lambda_0\}\rightarrow \{G_k,\Lambda_k\}$, being $G_0$ and $\Lambda_0$ the Newton's and the cosmological couplings respectively). As a consequence of this scale dependence, the domain of the classical solution is extended, resulting in semiclassical black holes solutions with modified properties, \normalsize where quantum effects are present, giving a more complete description of black holes. Some evidence has been reported supporting that the scaling behaviour resulting from this procedure is consistent with the Weinberg's Asymptotic Safety program \cite{wasp,wetterich,dou,souma,reuter,fischer,litim,percacci}. The effective action assuming running couplings has been analyzed in many situations \cite{koch, rincon,rincon2,rincon3,rincon4}. Within this framework, a regular black hole solution \cite{balart} has been interpreted in terms of scale-dependent couplings, without considering nonlinear electrodynamics \cite{contreras2}. This is a desirable feature, since nonlinear electrodynamics is not presently supported by experimental evidence. Another issue is that photons follow paths which corresponds to null geodesics of an effective geometry \cite{efqeo} produced by the self-interaction of the electromagnetic field, depending on the particular nonlinear theory adopted. For these reasons, it is of interest the study of scale-dependent black holes as gravitational lenses.


In this article, we study the gravitational lensing effects produced by the regular scale-dependent black hole obtained in Ref. \cite{contreras2}. The paper is organized as follows: In Sec. 2 we introduce the metric for the regular scale-dependent black hole, and we find the radius of the event horizon and the photon sphere in terms of the scale parameter of the theory; in Sec. 3 we find the deflection angle in the strong deflection limit; in Sec. 4, we obtain the positions and the magnifications of the relativistic images, and we analyze the observables for the case of the supermassive black hole in the center of our galaxy; finally, in Sec. 5 we summarize the results obtained.

\section{The geometry}
We first consider the effective Einstein-Hilbert action in the context of scale-dependent couplings,
\begin{equation}
\Gamma[g_{\mu\nu},k]=\int d^4 x \sqrt{-g}\left[\frac{1}{2\kappa_k}(R-2\Lambda_k)\right]+S_k,
\end{equation}
where $\kappa_k=8\pi G_k$ ($c=1$ is considered), and $G_k$ and $\Lambda_k$ are the gravitational and cosmological couplings respectively. The index $k$ appearing, takes into account the scale-dependent effect. The Einstein's field equations then result in the modified form
\begin{equation}
G_{\mu\nu}+\Lambda_k g_{\mu\nu}=\kappa_k (T^{\textrm{eff}})_{\mu\nu},
\label{einstein}
\end{equation}
being the effective energy-momentum tensor given by
\begin{equation}
(T^{\textrm{eff}})_{\mu\nu}:=(T_{\mu\nu})_k-\frac{1}{\kappa_k}\Delta t_{\mu\nu},
\end{equation}
with
\begin{equation}
\Delta t_{\mu\nu}=G_k(g_{\mu\nu}\,\square-\nabla_\mu \nabla_\nu)G^{-1}_{k}.
\end{equation}
As a solution of Eq. (\ref{einstein}) and assuming a spherically symmetric and static geometry, it can be finally obtained \cite{contreras2}
\begin{equation}
ds^{2}=-A(r)dt^{2}+B(r)dr^{2}+C(r)(d\theta ^{2}+\sin ^{2}\theta d\phi ^{2}),  \label{m1}
\end{equation}
with the metric functions given by
\begin{equation}
A(r)=B(r)^{-1}=1-\frac{2MG_0}{r}\left(1+\frac{M^2 G^{2}_{0}\epsilon}{6r}\right)^{-3}, \hspace{1cm} C(r)=r^2,
\label{metricelement}
\end{equation}
where $M$ is the mass, $G_0$ is the Newton's gravitational constant, and $\epsilon>0$ is a running parameter which has dimensions of inverse of length that is associated to the scale-dependent gravitational coupling $G_k$. The metric is asymptotically Schwarzschild, i.e. $A(r)\rightarrow 1-2G_0 M/r$ as $r\rightarrow\infty$ and is regular everywhere (nonsingular), having a de Sitter behaviour for $r\rightarrow 0$, since $A(r)\rightarrow 1-432r^2/G^{5}_{0}M^5\epsilon^3$. In the limit $r\rightarrow 0$, an effective cosmological constant $\Lambda_{\textrm{eff}}=1296/G^{5}_{0}M^{5}_{0}\epsilon^3$ is present. This regular solution was obtained without invoking to nonlinear electrodynamics and corresponds to a semiclassical extension of the Schwarzschild black hole. It can be interpreted without the cosmological term, by associating the matter content with an anisotropic vacuum which modifies the usual Schwarzschild geometry. This vacuum energy can be considered to be of quantum nature, and it is not negligible for non vanishing values of $\epsilon$, where the scale dependence effect becomes appreciable. In the limit $\epsilon\rightarrow 0$, the Schwarzschild solution is recovered. There are no constrains determined with certainty about the running parameter $\epsilon$, but since it comes from quantum corrections, it is expected to adopt small values.

\begin{figure}[t!]
\begin{center}
    \includegraphics[scale=0.94]{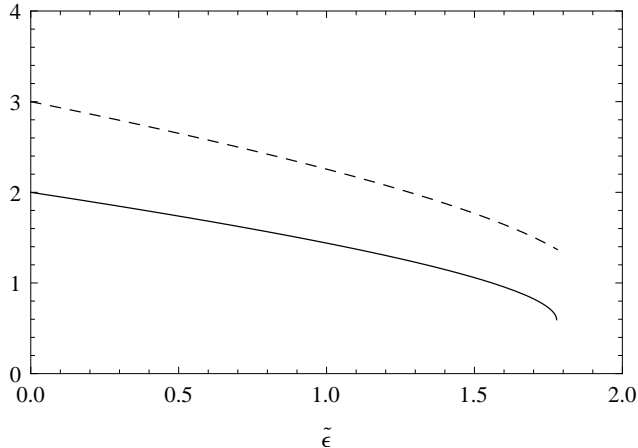}
    \vspace{0.25cm}
    \caption{The radius of the event horizon $x_h$ (solid line) and the radius of the photon sphere $x_{ps}$ (dashed-line) as functions of $\tilde{\epsilon}$.}
    \label{rhrps}
\end{center}
\end{figure}

The metric (\ref{m1}) can be conveniently written by defining the quantities $T=t/M$, $x=r/M$ and $\tilde{\epsilon}=\epsilon M$, and adopting $G_0=1$, results in
\begin{equation}
ds^{2}=-A(x)dT^{2}+B(x)dx^{2}+C(x)(d\theta ^{2}+\sin ^{2}\theta d\phi ^{2}),  \label{m2}
\end{equation}
where
\begin{equation}
A(x)=B(x)^{-1}=1-\frac{2}{x}\left(1+\frac{\tilde{\epsilon}}{6x}\right)^{-3}, \hspace{1cm} C(r)=x^2.
\label{metricelement2}
\end{equation}
The radius of the event horizon is obtained from equating $A(x)=0$, then it is given by the largest positive solution of the following expression:
\begin{equation}
216~x^{3} - \left(432 - 108 \tilde{\epsilon}\right)x^{2} + 18~\tilde{\epsilon}^2 x + \tilde{\epsilon}^3 = 0,
\label{xh}
\end{equation}
and corresponds to a third degree polynomial equation, with solution
\begin{equation}
x_h =\frac{1}{6} \Bigg\{4 - \tilde{\epsilon} + \frac{2^{8/3} (2 - \tilde{\epsilon})}{\left[32 - \tilde{\epsilon} \left(24 - 3 \tilde{\epsilon} - \sqrt{9  \tilde{\epsilon}^2-16 \tilde{\epsilon}}\right)\right]^{1/3}}+2^{1/3} \left[32 - \tilde{\epsilon} \left(24 - 3 \tilde{\epsilon} - \sqrt{9 \tilde{\epsilon}^2-16 \tilde{\epsilon}}\right)\right]^{1/3}\Bigg\}.
\label{xh2}
\end{equation}
The radius of the photon sphere $x_{ps}$ corresponds to the largest positive solution of the equation
\begin{equation}
D(x)=\frac{C^{\prime }(x)}{C(x)}-\frac{A^{\prime }(x)}{A(x)}=0,  \label{d}
\end{equation}
where the prime denotes the derivative with respect to the coordinate $x$. Replacing the metric functions (\ref{metricelement2}) and after some calculations, we find that, for the regular scale-dependent black hole, $x$ should satisfy
\begin{equation}
2592~x^{5} + (-7776 + 1728~ \tilde{\epsilon}) x^{4} + 432\tilde{\epsilon}^2 x^{3} + 48~\tilde{\epsilon}^3 x^{2} + 2~\tilde{\epsilon}^4 x  = 0.
\end{equation}
So, we obtain
\begin{equation}
x_{ps}=\frac{1}{12}\Bigg\{\sqrt{-\Omega - 8 \tilde{\epsilon}^2 + 2 (9 - 2 \tilde{\epsilon})^2 + \frac{54 \left[27 + 2 \tilde{\epsilon} (\tilde{\epsilon}-9)\right]}{\sqrt{
     \Omega+ 9 (9 - 4 \tilde{\epsilon})}}} +9 - 2 \tilde{\epsilon} + \sqrt{\Omega + 9 (9 - 4 \tilde{\epsilon})}\Bigg\},
\label{xps}
\end{equation}
with
\begin{equation}
\Omega=\frac{4\times 6^{2/3}
   \tilde{\epsilon}^3}{\left[-\tilde{\epsilon}^4 \left(-27 + \sqrt{
     729 - 384 \tilde{\epsilon}}\right)\right]^{1/3}} + 6^{1/3} \left[-\tilde{\epsilon}^4 \left(-27 + \sqrt{729 - 384 \tilde{\epsilon}}\right)\right]^{1/3}.
\end{equation}
From expressions (\ref{xh2}) and (\ref{xps}), we see that the existence of both an event horizon and a photon sphere ($x_{ps}>x_h$) is only possible for the running parameter lying in the range $0<\tilde{\epsilon}\leq 16/9\approx 1.78$. The maximum value of $\tilde{\epsilon}$ corresponds to the case of an extremal black hole, where Eq. (\ref{xh}) has a double root. From expression (\ref{xps}), we find that there is also a small range $1.78\leq\tilde{\epsilon}\leq243/128\approx 1.9$, where the geometry admits a photon sphere without the presence of an event horizon. As the metric (\ref{m2}) is regular, this case does not correspond to a naked singularity, but since the parameter $\tilde{\epsilon}$ is expected to be small, only the range $0<\tilde{\epsilon}<1.78$ will be considered of interest for the present work. In Fig. \ref{rhrps}, the radius of the event horizon (solid line) and the radius of the photon sphere (dashed-line) are shown; they are decreasing functions of the parameter $\hat{\epsilon}$. In the case $\tilde{\epsilon}=0$, the Schwarzschild values $x_h=2$ and $x_{ps}=3$ are recovered.

\section{Deflection angle}
The deflection angle for a photon coming from infinity, as a function of the adimensionalized closest approach distance $x_0$, can be written in the form \cite{weinberg,nakedsing1}
\begin{equation}
\alpha(x_0)=I(x_0)-\pi,  \label{alfa1}
\end{equation}
with
\begin{equation}
I(x_0)=2\int^{\infty}_{x_0}\frac{\sqrt{B(x)}}{\sqrt{R(x)C(x)}}dx,
\label{i0}
\end{equation}
in which
\begin{equation}
R(x)=\frac{A_0 C(x)}{A(x) C_0}-1,
\end{equation}
where, here and from now on, the subscript $0$ stands for evaluation in $x=x_0$ of the metric functions.

\begin{figure}
\begin{center}
    \includegraphics[scale=0.94]{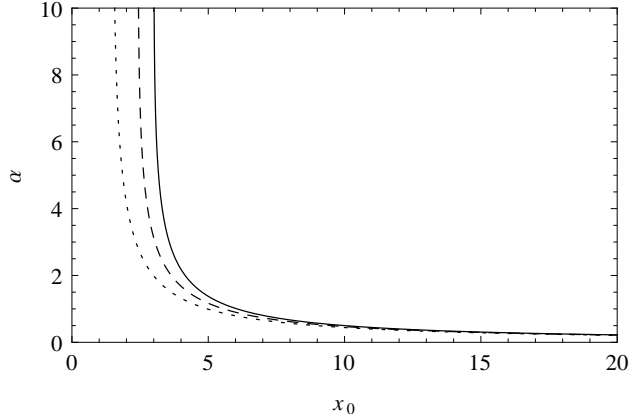}
    \vspace{0.25cm}
    \caption{Deflection angle for three representative values of the scale parameter $\tilde{\epsilon}$ as a function of the adimensionalized closest approach distance $x_0$: $\tilde{\epsilon}=0$ (solid line), $\tilde{\epsilon}=0.8$ (dashed-line) and $\tilde{\epsilon}=1.7$ (dotted-line).}
    \label{alphaexacto}
\end{center}
\end{figure}
\begin{figure}
\begin{center}
    \includegraphics[scale=0.94]{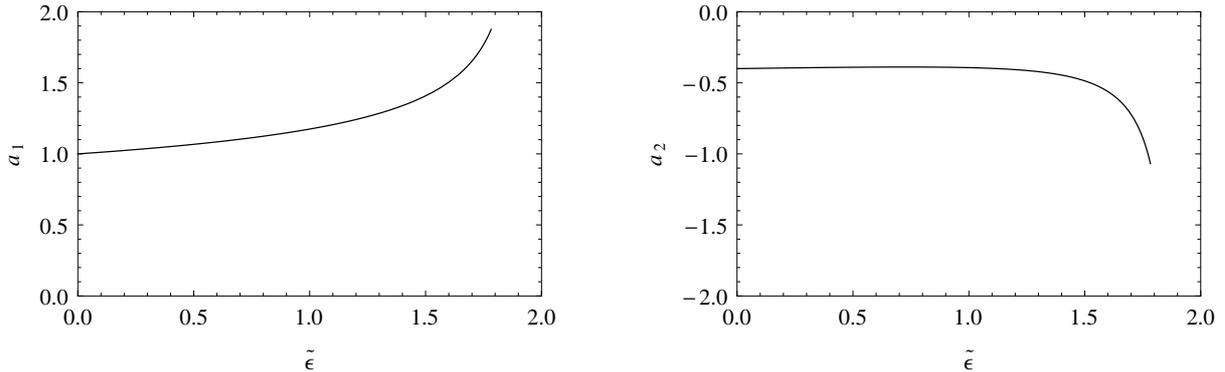}
    \vspace{-0.8cm}
    \caption{The strong deflection limit coefficients $a_1$ (left) and $a_2$ (right) as functions of $\tilde{\epsilon}$.}
    \label{sdlcoef}
\end{center}
\end{figure}

The deflection angle $\alpha$ has a monotonic decreasing behavior as a function of the closest approach distance $x_0$. This can be seen in Fig. \ref{alphaexacto}, which was obtained numerically for the regular scale-dependent black hole for three representative values of the parameter $\tilde{\epsilon}$. When $x_0$ approaches to the radius of the photon sphere, the deflection angle has a logarithmic divergence, and it tends to zero for large values of $x_0$. For photons passing close enough to the photon sphere, the deflection angle becomes larger than $2\pi$, which means that the photons perform one or more turns around the black hole before emerging from it and reaching a distant observer. As a consequence, two infinite sets of relativistic images are formed, one at each side of the black hole. This situation can be studied by performing the strong deflection limit, in order to find an analytic expression of the deflection angle. The integral (\ref{i0}) can be rewritten in a suitable way by following the procedure introduced in Refs. \cite{bozza02, tsukamoto}. By defining
\begin{equation}
z\equiv 1-\frac{x_0}{x},  \label{zvariable}
\end{equation}
so it takes the form
\begin{equation}
I(x_0)=\int^{1}_{0}f(z,x_0)dz,  \label{i02}
\end{equation}
where
\begin{equation}
f(z,x_0)=\frac{2x_0}{\sqrt{G(z,x_0)}},  \label{fdef}
\end{equation}
with
\begin{equation}
G(z,x_0)\equiv R(x)\frac{C(x)}{B(x)}(1-z)^4.  \label{gdef}
\end{equation}
It is useful to split the integral (\ref{i0}) into two parts
\begin{equation}
I(x_0)=I_D(x_0)+I_R(x_0),  \label{i03}
\end{equation}
with $I_D(x_0)$ the integral containing the divergence at $x_0=x_{ps}$ and $I_R(x_0)$ the term which is regular everywhere. The divergent part can be defined as
\begin{equation}
I_D(x_0)\equiv\int^{1}_{0}f_D(z,x_0)dz,  \label{id}
\end{equation}
where the argument $f_D(z,x_0)$ has the form
\begin{equation}
f_D(z,x_0)\equiv\frac{2x_0}{\sqrt{c_1(x_0)z+c_2(x_0)z^2}},  \label{fd}
\end{equation}
with
\begin{equation}
c_1(x_0)=\frac{C_0 D_0 x_0}{B_0}  \label{c1}
\end{equation}
and
\begin{equation}
c_2(x_0)=\frac{C_0 x_0}{B_0}\left\{D_0\left[\left(D_0-\frac{B^{\prime }_0}{B_0}\right)x_0-3\right]+\frac{x_0}{2}\left(\frac{C^{\prime \prime }_0}{C_0}-
\frac{A^{\prime \prime }_0}{A_0}\right)\right\}.  \label{c2}
\end{equation}
By taking the limit $x_0\rightarrow x_{ps}$, the strong deflection limit can be performed. In this situation, $D(x_{ps})=0$ from Eq. (\ref{d}), and expressions (\ref{c1}) and (\ref{c2}) simplify to
\begin{equation}
c_1(x_{ps})=0
\end{equation}
and
\begin{equation}
c_2(x_{ps})=\frac{C_{ps} x^{2}_{ps}}{2B_{ps}}\left(\frac{C^{\prime \prime }_{ps}}{C_{ps}} -\frac{A^{\prime \prime }_{ps}}{A_{ps}}\right),  \label{c2m}
\end{equation}
where the subscript ``$ps$'' in each funtion, denotes evaluation in $x=x_{ps}$. On the other hand, the integral $I_R$ in Eq. (\ref{i03}) is defined by
\begin{equation}
I_R(r_0)\equiv\int^{1}_{0}f_R(z,r_0)dz,  \label{ir}
\end{equation}
being
\begin{equation}
f_R(r_0)\equiv f(z,r_0)-f_D(z,r_0),  \label{fr}
\end{equation}
which results regular since the divergence corresponding to $x_0=x_{ps}$ is subtracted. Finally, the deflection angle in the strong deflection limit can be written as,
\begin{equation}
\alpha(u)=-a_1\ln\left(\frac{u}{u_{ps}}-1\right)+a_2+O((u-u_{ps})
\ln(u-u_{ps})),  \label{alpha}
\end{equation}
where
\begin{equation}
u=\sqrt{\frac{C_0}{A_0}},  \label{impactu1}
\end{equation}
is the impact parameter of the photon and $u_{ps}$ is the critical impact parameter, i.e., for photons with $x_0\rightarrow x_{ps}$. For the regular scale-dependent black hole, it results,
\begin{equation}
u_{ps}=x_{ps} \sqrt{\frac{(6 x_{ps} + \tilde{\epsilon})^3}{(6 x_{ps} + \tilde{\epsilon})^3 -
  432 x^{2}_{ps}}}.
\label{impact}
\end{equation}
The quantities $a_1$ and $a_2$ in expression (\ref{alpha}) are the so called strong deflection limit coefficients:
\begin{equation}
a_1=\sqrt{\frac{2B_{ps} A_{ps}}{C^{\prime \prime }_{ps} A_{ps}-C_{ps}A^{\prime \prime
}_{ps}}}  \label{abarra1}
\end{equation}
and
\begin{equation}
a_2=a_1\ln\left[x^{2}_{ps}\left(\frac{C^{\prime \prime }_{ps}}{C_{ps}}-
\frac{A^{\prime \prime }_{ps}}{A_{ps}}\right)\right]+I_R(x_{ps})-\pi,  \label{bbarra1}
\end{equation}
which only depend on the metric functions. Performing the above calculations for the regular scale-dependent black hole, these coefficients result
\begin{equation}
a_1=\sqrt{\frac{\left(6 x_{ps} + \tilde{\epsilon}\right)^5}{\left(6 x_{ps} + \tilde{\epsilon}\right)^5 - 15552~ x_{ps}^3 \tilde{\epsilon}}}
\label{a1scalar}
\end{equation}
and
\begin{equation}
a_2=a_1 \ln \left\{2 + \frac{864~ x^{2}_{ps} (36 x^{2}_{ps} - 24 x_{ps} \tilde{\epsilon} + \tilde{\epsilon}^2)}{\left(6 x_{ps} +  \tilde{\epsilon}\right)^2 \left[-432 x^{2}_{ps} + \left(6 x_{ps} + \tilde{\epsilon}\right)^3\right]}\right\}+I_R-\pi,
\label{a2scalar}
\end{equation}
where the integral $I_R$ can be obtained numerically for any value of $\tilde{\epsilon}$. In Fig. \ref{sdlcoef} we see that $a_1$ increases with the parameter $\tilde{\epsilon}$ and $a_2$ decreases as $\tilde{\epsilon}$ grows. Comparing these results with the Schwarzschild black hole solution, we have that $a_1>a^{Sch}_{1}=1$ and $a_2<a^{Sch}_{2}=\ln[216(7-4\sqrt{3})]-\pi\approx -0.400230$ for any $\tilde\epsilon>0$. The Schwarzschild case is recovered for $\tilde{\epsilon}\rightarrow 0$. By replacing the expressions (\ref{a1scalar}) and (\ref{a2scalar}) in Eq. (\ref{alpha}), the deflection angle is completely determined in the strong deflection limit, allowing this, the analytical calculation of the positions and the magnifications of the relativistic images.

\section{Relativistic images and observables}\label{images}

We consider the particular configuration where a point light source is situated behind the black hole acting as a lens. The line joining the lens and the observer defines the optical axis, and the distances observer-lens $D_{ol}$ and lens-source $D_{ls}$ are both assumed much greater than the horizon radius $x_h$. Consequently, the observer-source distance is given by $D_{os}=D_{ol}+D_{ls}$. Since the geometry is asymptotically flat and the deflection of photons takes place in the proximities of the black hole, the trajectories of photons can be approximated by straight lines. In this situation, the lens equation is given by \cite{lenseq}:
\begin{equation}
\tan \beta =\frac{D_{ol}\sin \theta -D_{ls}\sin (\alpha -\theta )}
{D_{os} \cos (\alpha -\theta )},  \label{pm1}
\end{equation}
where $\beta $ and $\theta$ are the angular position of the source and the angular position of an image detected by the observer respectively, both taken from the optical axis. The lensing effects become more relevant for highly aligned objects. In this case, $\beta$ and $\theta$ are small, and $\alpha$ is close to a multiple of $2\pi$. If $\beta\neq 0$, the primary and secondary images, and the two infinite set of point relativistic images are formed; but when $\beta=0$, the Einstein rings are present instead. For the first set of relativistic images, the deflection angle can be written as $\alpha=2n\pi+\Delta\alpha_n$, where $n\in \mathbb{N}$ and $0<\alpha_n\ll 1$. The lens equation (\ref{pm1}) then results in the simpler form \cite{bozza02,lenseq}
\begin{equation}
\beta =\theta - \frac{D_{ls}}{D_{os}}\Delta \alpha _{n}. \label{pm2}
\end{equation}
For the other set of relativistic images, $\Delta\alpha_n$ in Eq. (\ref{pm2}) should be replaced by $-\Delta\alpha_n$ since $\alpha=-2\pi-\Delta\alpha_n$. The impact parameter can be approximated by $u=D_{ol}\sin\theta\approx D_{ol}\theta$, so the deflection angle (\ref{alpha}) is given by
\begin{equation}
\alpha (\theta ) = -a_1\ln \left( \frac{D_{ol}\theta }{u_{m}}
-1\right) +a_2.  \label{pm4}
\end{equation}
By inverting Eq. (\ref{pm4}), $\theta$ as a function of $\alpha$ is obtained. Performing a first order Taylor expansion around $\alpha =2n\pi $, the angular position of the $n$-th image for the first set of relativistic images results
\begin{equation}
\theta _{n}=\theta _{n}^{0}-\zeta _{n}\Delta \alpha _{n},  \label{pm6}
\end{equation}
with
\begin{equation}
\theta _{n}^{0}=\frac{u_{m}}{D_{ol}}\left[ 1+e^{(a_2-2n\pi )/a_1}
\right]  \label{pm7}
\end{equation}
and
\begin{equation}
\zeta _{n}=\frac{u_{m}}{a_1 D_{ol}}e^{(a_2-2n\pi )/a_1}.
\label{pm8}
\end{equation}
From Eq. (\ref{pm2}), we have that $\Delta \alpha _{n}=(\theta _{n}-\beta )D_{os}/D_{ls}$, and replacing this expression in Eq. (\ref{pm6}),
\begin{equation}
\theta _{n}=\theta _{n}^{0}-\frac{\zeta _{n}D_{os}}{D_{ls}}(\theta
_{n}-\beta ),  \label{pm10}
\end{equation}
which, by using $0<\zeta _{n}D_{os}/D_{ls}<1$ and keeping only the first-order term in $\zeta _{n}D_{os}/D_{ls}$, the angular positions of the images finally take the form
\begin{equation}
\theta _{n}=\theta _{n}^{0}+\frac{\zeta _{n}D_{os}}{D_{ls}}(\beta -\theta_{n}^{0}).  \label{pm14}
\end{equation}
Analogously, for the other set of the relativistic images, we obtain
\begin{equation}
\theta _{n}=-\theta _{n}^{0}+\frac{\zeta _{n}D_{os}}{D_{ls}}(\beta +\theta
_{n}^{0}).  \label{pm15}
\end{equation}

The magnification of the $n$-th relativistic image is given by the quotient of the solid angles subtended by the image and the source
\begin{equation}
\mu_n=\left|\frac{\sin\beta}{\sin\theta_n}\frac{d\beta}{d\theta_n}
\right|^{-1},  \label{mag1}
\end{equation}
which, from Eq. (\ref{pm14}) and considering small angles, we find
\begin{equation}
\mu _{n}=\frac{1}{\beta}\left[ \theta ^{0}_{n}+ \frac {\zeta _{n}D_{os}}{
D_{ls}}(\beta - \theta ^{0}_{n})\right] \frac {\zeta _{n}D_{os}}{D_{ls}}.
\label{pm18}
\end{equation}
Finally, by performing a first order Taylor expansion in $\zeta_n D_{os}/D_{ls}$, we have that the magnification of the $n$-th relativistic image, for both sets of images is given by
\begin{equation}
\mu _{n}=\frac{1}{\beta}\frac{\theta ^{0}_{n}\zeta _{n}D_{os}}{D_{ls}}.
\label{pm19}
\end{equation}
The magnifications decrease exponentially with $n$, so the first relativistic image is the brightest one. The rest of the images are very faint unless $\beta$ is close to zero, because they are proportional to $(u_{m}/D_{ol})^2$.

Since the first image is the outermost and the brightest one, the following observables can be conveniently defined \cite{bozza02}:
\begin{equation}
\theta_{\infty}=\frac{u_{m}}{D_{ol}},  \label{ob1}
\end{equation}
\begin{equation}
s=\theta_1-\theta_{\infty},  \label{ob2}
\end{equation}
and
\begin{equation}
r=\frac{\mu_1}{\sum^{\infty}_{n=2}\mu_n}.  \label{ob3}
\end{equation}
\begin{figure}
\begin{center}
    \hspace{-1cm}\includegraphics[scale=0.99]{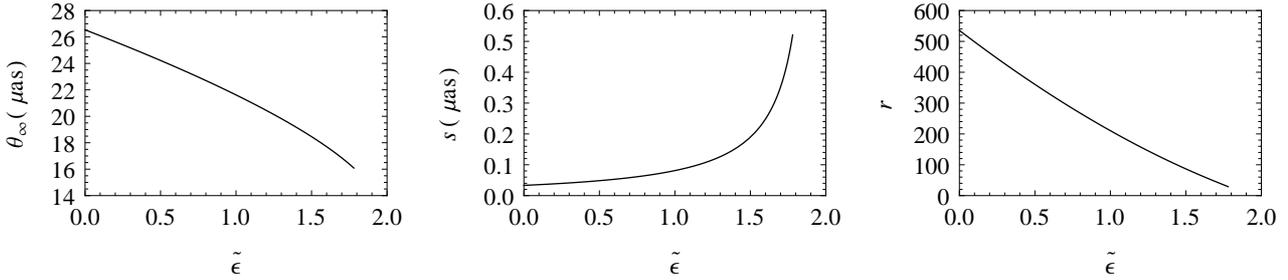}
    \vspace{-0.8cm}
    \caption{Observables $\theta_\infty$ (left), $s$ (center) and $r$ (right) as functions of the parameter $\tilde{\epsilon}$, for the Milky Way supermassive black hole ($M=4.31\times 10^6 M_{\odot}$ and $D_{ol}=8.33$ kpc).}
    \label{observables}
\end{center}
\end{figure}
\noindent The observable $s$ represents the angular separation between the angular position of the first relativistic image and the limiting value of the others $\theta_{\infty}$, and the quantity $r$ corresponds to the quotient between the flux of the first image and the sum of the fluxes coming from all the other images. These observables, in the strong deflection limit and for a high alignment configuration, take the simple form \cite{bozza02}:
\begin{equation}
s=\theta_{\infty}e^{(a_2-2\pi)/a_1}  \label{ob4}
\end{equation}
and
\begin{equation}
r=e^{2\pi/a_1},  \label{ob5}
\end{equation}
which depend on the geometry of the black hole, since they are functions of the strong deflection limit coefficients. So, by measuring $\theta_{\infty}$, $s$ and $r$, one can invert Eqs. (\ref{ob4}) and (\ref{ob5}) to obtain the strong deflection limit coefficients $a_1$ and $a_2$ from observations, and compare them with those predicted by the theoretical models to identify the nature of the black hole lens.

As a numerical example, the observables for the supermassive galactic center black hole are presented in Fig. \ref{observables}, for which the mass is $M=4.31\times 10^6 M_{\odot}$ and the distance from Earth is $D_{ol}=8.33$ kpc \cite{gillessen17}. As $\tilde{\epsilon}$ grows, the limiting value $\theta_{\infty}$ decreases, so the images get closer to the black hole. The separation between the first and the rest of the relativistic images increases with $\tilde{\epsilon}$ and the first relativistic image results less bright with respect to the others as $\tilde{\epsilon}$ increases. Comparing with the Schwarzchild black hole (i.e., when $\tilde{\epsilon}\rightarrow 0$), we have that $16.0558~\mu$as $<\theta_{\infty}<\theta^{Sch}_{\infty}= 26.5093~\mu$as, the angular separation $s$ lies in the range $0.5300~\mu$as $>s>s^{Sch}=0.0332~\mu$as, and $28<r<r^{Sch}\approx 535$. For large values of $\tilde{\epsilon}$ (if physically possible, i.e. quantum effect) the differences are noticeable, and in this case, the regular scale-dependent black hole could eventually be distinguished from the Schwarzschild one.

\section{Final remarks}\label{discu}
In this paper, we have studied the gravitational lensing effects produced by regular scale-dependent black holes introduced in Ref. \cite{contreras2}, which become interesting since they were obtained without invoking to nonlinear electrodynamics as a matter source. These black holes are asymptotically flat and are characterized by the mass $M$ and the running parameter $\epsilon$, appearing as a consequence of the scaling of the theory, which, in principle, is expected to be small. In particular, we have analyzed the situation where photons pass close to the photon sphere, which gives place to the formation of the relativistic images. We have performed the strong deflection limit to obtain the deflection angle, and the positions and the magnifications of the relativistic images in terms of the running parameter $\epsilon$. We have also calculated the observables and applied these results to the case of the supermassive galactic center black hole in order to compare them to the ones obtained for the Schwarzschild spacetime (i.e. $\epsilon\rightarrow 0$). For the regular scale-dependent black hole, we have seen that the strong deflection limit coefficients differ from the Schwarzschild ones: $1.88>a_1(\tilde{\epsilon})>a^{Sch}_{1}=1$ and $-1.0682<a_2(\tilde{\epsilon})<a^{Sch}_{2}=\ln[216(7-4\sqrt{3})]-\pi\approx -0.4002$. On the other hand, we have found that for larger values of $\epsilon$ the first relativistic image lies more separated from the others (packed together at $\theta_{\infty}$), which get closer to the black hole as $\epsilon$ grows. The relative magnification of the first relativistic image with respect to all the others is smaller $r<r^{Sch}$ for any value of $\epsilon$, which means that the first relativistic image becomes less bright in comparison to the others.

The relativistic images and the shadow of black holes correspond to a full relativistic description of their near horizon region. This is why, nowadays there is much attention paid to the observation of the vicinity of black holes, in particular, the supermassive ones located at the center of nearby galaxies, included ours (Sgr A*), due to its proximity and large size. In order to fullfil this purpose, many projects are now operational. The Event Horizon Telescope is one of them, which combines existing radio facilities resulting in a telescope with an angular resolution of about 20 $\mu$as by using very long baseline interferometry. This telescope has already made observations of the center of our galaxy and the nearby giant galaxy M87 \cite{broderick15}, and the first picture of the shadow of a black hole was reported recently \cite{eth1,eth2,eth3,eth4,eth5,eth6}. GRAVITY is a second generation instrument on the Very Large Telescope Interferometer (VLTI) that will examine in the near-infrared band the vicinity of the supermassive black hole in the center of our galaxy \cite{gillessen17}. Millimetron is a space-based mission with expected resolution of 0.05 µas, operating from far infrared to millimeter wavelengths. More details of this topic can be found in Refs. \cite{falcke} and the references therein. Gravitational lensing would be useful to compare different black hole models with observations. However, the subtle effects that arise among them, seem to need a second generation of future instruments.

\section*{Acknowledgments}

This work has been supported by Universidad de Buenos Aires and by CONICET. Particularly, I would like to thank Dr. Ernesto F. Eiroa for his useful comments.

\end{document}